\documentstyle[aps,twocolumn,pre]{revtex}
\begin{document}
\draft
\twocolumn[\hsize\textwidth\columnwidth\hsize\csname
@twocolumnfalse\endcsname

%
%                                      Change from IEEE to ReVTeX Style
%
%\documentstyle[twocolumn]{IEEEtran}
%\documentstyle[12pt,twoside,draft]{IEEEtran}
%\documentstyle[9pt,twocolumn,technote,twoside]{IEEEtran}

%\def\BibTeX{{\rm B\kern-.05em{\sc i\kern-.025em b}\kern-.08em
%    T\kern-.1667em\lower.7ex\hbox{E}\kern-.125emX}}

%\newtheorem{theorem}{Theorem}
%\setcounter{page}{100}

%\begin{document}

\title{Simulation of Magnetization Switching in Biaxial 
        Single-Domain Ferromagnetic Particles}

\author{Xuekun Kou$^{1,2}$, 
        M.~A.~Novotny$^{1,2}$, and 
        Per~Arne~Rikvold$^{1,3}$}

\address{
$^1$Supercomputer Computations Research Institute,
    Florida State University,
    Tallahassee, Florida 32306-4130 \\
$^2$Dept.\ of Electrical Engineering,
    FAMU--FSU College of Engineering,
    Tallahassee, FL 32310-2870\\
$^3$Center for Materials Research and Technology, and
    Department of Physics,
    Florida State University,
    Tallahassee, Florida 32306-4350
}

%\markboth{IEEE Transactions On Magnetics, Vol. x, No. y,  March 1998}
%{Kou, Novotny and Rikvold: Monte Carlo Simulation\ldots}

\date{\today}
\maketitle

\begin{abstract}
The magnetization switching dynamics of biaxial single-domain homogeneous
ferromagnetic particles, in which the two easy axes are perpendicular 
to each other, is simulated using a 4-state clock model. A zero-field mapping 
of the statics between the symmetric 4-state clock model and two decoupled 
Ising models is extended to non-zero field statics and to the dynamics. 
This significantly simplifies the analysis of the 
simulation results. We measure the magnetization switching time of 
the model and analyze the results using droplet theory. The switching 
dynamics in the asymmetric model is more complicated. If the easy axis 
is perpendicular to the stable magnetization direction, the system can 
switch its magnetization via two different channels, one very fast and 
the other very slow. A maximum value for the switching field as a function of 
system size is obtained. The asymmetry affects the switching fields
differently, depending on whether the switching involves one single droplet
or many droplets of spins in the stable magnetization configuration. 
The angular dependence of the switching field in symmetric and asymmetric 
models is also studied.

\end{abstract}

 % Probability theory, stochastic processes, and statistics
\vskip1pc]

%\begin{keywords}
%Magnetization switching, biaxial media, 4-state clock model, 
%                         Monte Carlo simulation
%\end{keywords}

\section{Introduction}
\label{intro}

Biaxial media have recently attracted much research 
interest as potential materials for applications in high-density 
magnetic recording. This is a consequence of their good thermal stability, 
high coercivity and coercivity squareness, and low medium 
noise~\cite{YAO96}-\cite{ZHU94}. With the current rapid increase in recording
density, the size of a magnetic particle used to store a single bit may
soon become so small that it can only contain one single magnetic 
domain. Understanding the dynamics of magnetization switching in individual 
single-domain particles is therefore essential. Biaxial media may be obtained,
for example, by sputtering 
a Co-based alloy onto a Cr substrate~\cite{WONG91,MIRZA}, by fabricating
perovskite superlattices consisting of different ferromagnetic metallic
oxides~\cite{GONG96}, or by epitaxially growing Fe on GaAs(001)~\cite{EBEL97}. 
By applying lithographic patterning techniques to precisely control 
the particle shape, a uniaxial medium may also exhibit biaxial 
properties due to competition between the magneto-crystalline and shape 
anisotropies~\cite{NEW95,NEW96}. The two easy axes in biaxial media are 
often perpendicular to each other, and they may have the same 
or different magnetic anisotropies. In thin films, these two axes can both be
in-plane, or one of them can be out-of-plane, often perpendicular to 
the plane of the film. New techniques, such as 
nanolithography and ultra-high resolution scanning microscopies, 
have recently made the synthesis and experimental observation 
of isolated single-domain particles possible. For instance, 
the magnetization switching of individual barium ferrite nanoscale 
particles has been observed using Magnetic Force Microscopy (MFM), and a 
maximum switching field was observed for particles of diameter near 
55~nm~\cite{CHANG93}. For other particles, the angular dependence of the 
switching (or coercive) field has been measured~\cite{WERN95,LEDER94}, 
the switching dynamics investigated~\cite{WERN95}, and the switching 
statistics measured~\cite{LEDER94}. In 
Refs.~\cite{FERN96} and~\cite{WEISS96}, arrays of
single-domain Co dots were fabricated, and the magnetization switching was 
observed by MFM. Depending on the shape of the Co dot, both in-plane and 
out-of-plane magnetic moments were observed. 

A powerful non-perturbative theoretical method to study models of
magnetization switching dynamics is Monte Carlo simulation. One standard 
model used to study uniaxial 
magnets is the kinetic Ising model~\cite{RICH95,MELIN96}. However, 
the Ising model can only be used as a reasonably realistic representation 
of extremely anisotropic uniaxial magnets, and a different model is required 
to capture the unique properties of biaxial materials. In this work 
we study a 4-state clock model with two perpendicular axes.
In the most symmetric case there is no applied field and four states are equivalent. Suzuki has found a mapping between this symmetric 4-state clock model and two superimposed Ising models~\cite{MASUO}. When a field is applied, this symmetry is lost and o
the results of extensive simulations of the switching dynamics. In order 
to simplify the analysis of the simulation results, we extend 
Suzuki's mapping to non-zero applied field 
and to the dynamics. 

The rest of this paper is organized as follows. In 
Sec.~\ref{sec:model} we describe the model used in the simulations. 
In Sec.~\ref{sec:drop} we briefly review the aspects of droplet theory used 
to analyze the numerical results. In Sec.~\ref{sec:life} and 
Sec.~\ref{sec:switch} we
present and analyze our numerical results for lifetime and switching field, 
respectively. Finally, conclusions are drawn in Sec.~\ref{sec:summ}.

\section{Model Description}
\label{sec:model}

\begin{figure}
\begin{center}
	\vspace*{1.2in}
	\includegraphics{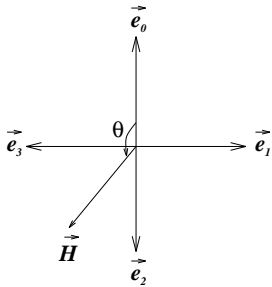} 	
\end{center}
\caption{Vector representation of the 4-state clock model.}
\label{fig:clk}
\end{figure}

The 4-state clock model is defined by the microscopic Hamiltonian
\begin{equation}
  \label{eq:CLKHM}
        {\cal H} = -J\sum_{\langle i,j \rangle}\vec \mu_i \cdot \vec \mu_j 
                - \vec H \cdot \sum_i \vec \mu_i 
                - A\sum_i (\vec \mu_i \cdot \vec e_0)^2 \; .
\end{equation}
The unit vector $\vec \mu_i$ is the orientation of 
the spin at site $i$, which can be in one of 4 directions: ${\vec e_0}$, 
${\vec e_1}$, ${\vec e_2}$, and ${\vec e_3}$, as shown in Fig.~\ref{fig:clk}.
The angle between the applied magnetic field $\vec H$ and $\vec e_0$ is 
$\theta$. The dimensionless system magnetization is given by
\begin{equation}
  \label{eq:sysmag}
	\vec m = L^{-2} \sum_i \vec \mu_i\; .
\end{equation}
We consider a two-dimensional square lattice with $L$ spins on each side, 
and periodic boundary conditions are applied in both directions to suppress
boundary effects. (Boundary effects for the Ising model are discussed in
Ref.~\cite{RICH97}.) The lattice 
constant is set to unity. In Eq.~(\ref{eq:CLKHM}), $\sum_i$ runs over all 
$L^2$ sites, and $\sum_{\langle i,j \rangle}$ runs over all nearest-neighbor 
pairs on the lattice. 
The first term in Eq.~(\ref{eq:CLKHM}) is the energy due to exchange 
interactions between nearest-neighbor spins, and $J > 0$ is the ferromagnetic 
exchange interaction. The second term is the energy, $L^2 \vec H \cdot \vec m$ 
due to the applied field $\vec H$. The third term represents the energy due to
a different preference for one of the two axes, $\vec e_0$--$\vec e_2$
and $\vec e_1$--$\vec e_3$. If $A=0$, the model has 4-fold symmetry for 
${\vec H} = 0$, and the statics of this model is equivalent to two decoupled 
Ising models~\cite{MASUO}. In 
Appendix A, this mapping is extended to non-zero fields and to the
dynamics. A nonzero value of $A$ destroys this 4-fold symmetry. Therefore, 
throughout this paper we refer to the model with $A \neq 0$ as 
{\it asymmetric}. (In the statistical mechanics literature, clock models
with $A \neq 0$ are often referred to as ``anisotropic.'' However, in this 
paper we use the terms ``anisotropy'' and ``anisotropic'' as they are
customarily understood in the literature on magnetic materials.) 
For $A > 0$ and ${\vec H} = 0$, the $\vec e_0$--$\vec e_2$ axis is favorable 
energetically, while the $\vec e_1$--$\vec e_3$ axis is preferred for $A < 0$. 

The critical temperature, $T_{\rm c}$, of the model with $A=0$ and $\vec H=0$
is known exactly. Its value is 
$k_{\rm B}T_{\rm c}/J= 1/\ln(1+\sqrt{2}) \approx 1.135$~\cite{MASUO}, exactly 
that of a two-dimensional square-lattice Ising model with interaction
constant $J/2$. Here $k_{\rm B}$ is Boltzmann's constant.
The simulation temperature is chosen to be $T = 0.8T_{\rm c}$. 
This simulation temperature is high enough to get a reasonable 
Monte Carlo acceptance rate, while it is low enough to avoid complications 
near the critical point such as critical slowing-down and finite-size effects 
~\cite{KKMON}. Based on studies of the 
kinetic Ising model~\cite{RICH95,RIKV94r,RIKV98}, simulation 
results at other moderately low temperatures should be 
qualitatively similar to those obtained here. 

The initial configuration of the system is ordered, with 
$\vec \mu_i = \vec e_0$ for all spins. This may be achieved by applying a 
strong field in the $\vec e_0$ direction for time $t<0$. Then at $t=0$
the external magnetic field is changed to 
$\vec H$~(Fig.~\ref{fig:clk}). During the simulation, first 
one spin is selected at random, then a new candidate state is 
chosen. At low temperatures, large orientational changes of the spin 
($\vec e_0 \leftrightarrow \vec e_2$ and $\vec e_1 \leftrightarrow \vec e_3$)
are very unlikely~\cite{KKSKI}. Consequently, the candidate state is obtained 
by rotating the spin by $90^\circ$ or $-90^\circ$. The {\it a priori}
probabilities of clockwise and anti-clockwise rotations are each set to $1/2$. 

The probability that the spin flips to the chosen state is given by
Glauber dynamics~\cite{GLAUB}
\begin{equation}
  \label{PFLIP}
        W = \frac{\exp(-\beta \Delta E)}{1+\exp(-\beta \Delta E)}\; ,
\end{equation}
where $\Delta E$ gives the change in the energy of the system that would 
result if the spin flip were accepted, and $\beta = 1/k_{\rm B} T$.
The Glauber dynamic is chosen because it can be derived from quantum mechanics 
under certain stringent conditions~\cite{MARTIN}. Here time is reported in
the unit of one Monte Carlo steps per spin~(MCSS). 

The lifetime is defined as the average time, 
$\langle \tau \rangle$, needed to satisfy a specific stopping criterion 
(discussed below), where $\tau$ is the first-passage time to the stopping 
criterion for an individual Monte Carlo run. The average is 
taken over at least 1000 independent Monte Carlo runs in order to 
obtain good statistics. In addition to the mean, we also 
measure the relative standard deviation
\begin{equation}
	\label{eq:littler}
	r = \sqrt{\langle \tau^2 \rangle - \langle \tau \rangle^2} / 
	\langle \tau \rangle\; .
\end{equation}

The switching field is measured by observing whether on average the system 
magnetization has reversed after a fixed waiting time, here chosen as 
$t_{\rm w} = 200$~MCSS. As Ising model studies have 
demonstrated~\cite{RICH95,RIKV94r}, the switching fields 
for different waiting times are qualitatively similar. 

To measure the lifetime of a metastable state, a criterion for 
escape from the metastable state must be chosen.  
In studies of the kinetic Ising model, the usual criterion is that an 
escape is registered and the simulation stopped the first 
time the magnetization reaches a fixed cutoff value, $m_{\rm stop}$.  
(In other words: the stopping time is the first-passage time to 
$m_{\rm stop}$.) Results for different values of $m_{\rm stop}$ 
are significantly different only in the deterministic regime, and even in 
this regime the results are qualitatively similar \cite{RIKV94}.  

In some experimental studies of magnetization switching, the sign of the 
magnetization along a particular direction is much easier to measure than its 
magnitude~\cite{CHANG93,WERN95}. 
The stopping criterion most directly
relevant to these experiments is therefore to use $m_{\rm stop}=0$.  
For the Ising model, this is obtained by stopping the simulation 
when half of the spins have entered the stable state (and half have 
left the metastable state). For other experimental situations, a different 
criterion might be easier to measure. However the stopping criterion chosen 
should not have a large effect on the results. 

In order to completely map the 
the 4-state clock model onto two superimposed Ising systems (see 
Appendix A), the stopping criterion for the 4-state clock model 
must map onto the stopping criteria for the two 
individual Ising systems.  
If the stopping criteria for both Ising systems are 
given by $m_{\rm stop}=0$, two different stopping criteria 
for the 4-state clock model can be naturally envisioned.  
Denoting these by $S_{\cup}$ and $S_{\cap}$, they are 
given in terms of the first passage 
to the stopping parameter $\mu_{\rm stop}$ by 
\begin{eqnarray}
  \label{stop1}
S_{\cup} \>\> {\rm has} & \mu_{\rm stop} = m_{I0} m_{I1} \\
S_{\cap} \>\> {\rm has} & \mu_{\rm stop} = 1-\eta(-m_{I0})\eta(-m_{I1}) \;,
\end{eqnarray}
where each $m_{I\alpha}$ for $\alpha=0,1$ is the time-dependent 
magnetization of one of the Ising systems 
and $\eta(x)$ is the unit step function.  
The criterion $S_{\cup}$ is to stop when 
the first Ising system reaches zero magnetization, 
while for $S_{\cap}$ the simulation is continued 
until the last of the two Ising systems reaches zero magnetization.

The mapping between the clock model and the two Ising models illustrates 
that in order for a clock spin to escape from the initial state, only one of 
the two Ising spins on the site needs to flip, while both Ising spins on the 
same site must flip in order for the clock spin to enter the stable state.
Therefore, for the clock model the 
$S_{\cup}$ stopping criterion implies that 
$\ge \frac{1}{2}$ of the clock spins have escaped 
from the {\it initial\/} state.  
The $S_{\cap}$ stopping criterion 
means a certain fraction ($\leq \frac{1}{2}$) of the clock spins 
have entered the {\it stable\/} state.  For the symmetric clock model, 
the two Ising models are 
decoupled, and their switching processes are independent of each other.
Therefore, for $S_{\cap}$ 
on average $1/4$ of the clock spins will have entered the stable 
state at the time the simulation is stopped. In this paper, 
we use only the $S_{\cap}$ stopping criterion since it more 
closely models an experimental stopping criterion of zero magnetization.  
For the asymmetric model, we also use this stopping criterion in order 
to observe the effects of $A$ and $\vec H$ consistently. 
In this case the average 
fraction of clock spins in the stable state at the end of 
the simulation
is between $1/4$ and $1/2$, depending on $A$. 
A more detailed discussion of the quantitative differences between 
different stopping criteria is given in Ref.~\cite{KOU97}.

\section{Droplet Theory}
\label{sec:drop}

During magnetization switching in single-domain particles, small 
``droplets'' of the stable phase are continually created 
and destroyed by thermal fluctuations~\cite{RICH95}. The fate of a droplet is 
determined by the competition between its total surface free energy, which is 
proportional to its surface area, and the coupling of its magnetization 
with the applied field, which is proportional to its volume. If 
the droplet is small, further growth produces an increase of the free 
energy due to the surface tension. Therefore, the droplet will be destroyed 
with high probability. If the droplet is sufficiently large, this 
free-energy  penalty is offset by the benefit obtained from increasing the 
number of spins parallel to the applied magnetic field. Therefore the 
droplet is very likely to grow further. For a droplet with a critical
radius $R_c$, the tendencies towards growth and shrinkage balance, and the 
droplet will grow or shrink with equal probability. By comparing $R_c$ with
the system size $L$, the lattice spacing $a$, and the average distance 
between supercritical droplets $R_0$, the magnetization 
switching dynamics is divided into four 
regimes~\cite{RICH95,RIKV94r,RIKV94,TOMITA92}:
\begin{itemize}
  \item The Coexistence~(CE) regime $[a<L<R_c]$, in which the switching is 
	governed by subcritical fluctuations on the scale of the 
	system size.
  \item The Single-droplet~(SD) regime $[a<R_c<L<R_0]$, characterized by 
	switching via a single critical droplet. Once a supercritical
        droplet is nucleated, it will most likely grow to occupy
	the entire system before another droplet is nucleated.
  \item The Multi-droplet~(MD) regime $[a<R_c<R_0<L]$, characterized by 
	switching via a finite density of critical droplets.
  \item The Strong-field~(SF) regime $[a \simeq R_c \simeq R_0 < L]$, in which 
	the concept of a droplet is no longer applicable.
\end{itemize}

Since the symmetric 4-state clock model is equivalent to two decoupled
Ising models, we can simplify the analysis of the clock model based on
extensive studies
of kinetic Ising models~\cite{RICH95,RIKV94r,TOMITA92,RIKV94}. Here a brief
review of the droplet theory for the Ising model is given. For more details, 
see the references listed above.

In the SD regime, the lifetime is given by
\begin{equation}
\label{eq:sdtime}
        \tau_{\rm I} \approx [ L^d \Gamma(T, H_{\rm I}) ]^{-1}\;,
\end{equation}
where $H_{\rm I}$ is the field in the Ising Hamiltonian, and 
the nucleation rate per unit volume $\Gamma(T,H_{\rm I})$ 
is~\cite{RIKV94r}
\begin{equation}
\label{eq:nurate}
        \Gamma(T,H_{\rm I}) \approx  A(T)|H_{\rm I}|^K
            \exp \left[-\frac{\beta \Xi(T)}{|H_{\rm I}|^{d-1}} 
	(1+O(H_{\rm I}^2))\right]\; .
\end{equation}
Here $A(T)$ is a non-universal prefactor, and $K$ is believed to be 3 for two
dimensional ($d=2$) Ising systems~\cite{RIKV94r}. 
At $T = 0.8T_c$, $\Xi(T)/J_{\rm I}^2 \approx 0.92$~\cite{RIKV94r}, where 
$J_{\rm I}$ is the exchange interaction in the Ising Hamiltonian.

In the MD regime, the system can be approximately partitioned 
into $(L/R_0)^d \gg 1$ cells, each of which decays via an independent 
Poisson process with rate $R_0^d \Gamma$. Therefore, the volume 
fraction is self-averaging, and the average lifetime 
$\langle \tau_{\rm I} \rangle$, defined as the average first-passage 
time to zero magnetization, is given by the well-known Avrami theory
of metastable decay as~\cite{RICH95,RIKV94}
\begin{equation}
\label{eq:mdtime}
      \langle\tau_{\rm I}\rangle \propto \left[|H_{\rm I}|^d 
	\Gamma(T,H_{\rm I}) \right]^{-1/(d+1)}\; .
\end{equation}

By taking the derivative of the logarithm of $\langle \tau_{\rm I} \rangle$ 
with respect to $|J_{\rm I}/H_{\rm I}|^{d-1}$, one obtains
\begin{equation}
     \label{eq:lamda}
        \frac{\Lambda_{\rm{Ieff}}}{J_{\rm I}^{d-1}} \equiv 
	\frac{d\ln \langle \tau_{\rm I} \rangle}
        {d(|J_{\rm I}/H_{\rm I}|^{d-1})} 
	= \lambda |H_{\rm I}/J_{\rm I}|^{d-1} + 
	\frac{\Lambda}{J_{\rm I}^{d-1}}\; ,
\end{equation}
with $\lambda = K/(d-1)$ and $\Lambda = \beta \Xi (T)$ in 
the SD regime, while $\lambda = (K+d)/(d^2-1)$ and $\Lambda = 
\beta \Xi(T)/(d+1)$ in the MD regime~\cite{RIKV94r,RIKV94}. 

For the symmetric 4-state clock model, if the applied field is along 
the $\vec e_2$ direction, the applied field $H_{\rm I}$ and the 
exchange interaction $J_{\rm I}$ of the Ising model are related to the
corresponding parameters for the clock model by $H_{\rm I} = H/2$ and  
$J_{\rm I} = J/2$~(Table~\ref{mapping}). By replacing the parameters for 
the Ising model by the corresponding ones for the clock model in the above 
analysis, we get the dimensionless value $\Lambda_{\rm{eff}}/J^{d-1}$ for
the clock model, which is
\begin{equation}
     \label{eq:lamdaclock}
       \frac{\Lambda_{\rm{eff}}}{J^{d-1}} =  
         \frac{d\ln \langle \tau \rangle} {d(|J/H|^{d-1})} = 
       \lambda |H/J|^{d-1} 
	+ \frac{\Lambda}{(J/2)^{d-1}}\; ,
\end{equation}
Since $J_{\rm I} = 1/2J$, comparing  Eq.~(\ref{eq:lamda}) and
Eq.~(\ref{eq:lamdaclock}), we have
\begin{equation}
\label{equalIC}
	  \Lambda_{\rm{eff}}/J^{d-1} 
	= \Lambda_{\rm{Ieff}}/J_{\rm I}^{d-1}\; .
\end{equation}
Therefore we can compare the simulation results for the symmetric 4-state 
clock model directly with the theoretical predictions for the Ising models. 
For the asymmetric clock model, we calculate the same derivative.

\section{Lifetime Measurements}
\label{sec:life}

The statistical properties of the lifetime depend on the switching regime of 
the system. For an Ising 
system in the SD regime, the probability density of the lifetime 
is exponential~\cite{RICH95,RIKV94r,RIKV94,TOMITA92} 
\begin{equation}
\label{eq:islt}
        p_{\rm I}(\tau) = \langle \tau_{\rm I} \rangle^{-1} 
                \exp(-\tau / \langle \tau_{\rm I} \rangle)\; ,
\end{equation}
where $\langle \tau_{\rm I} \rangle$ is the mean lifetime. 
As discussed in Sec.~\ref{sec:model}, for the 4-state clock model, 
magnetization switching, 
as defined by the stopping criterion used in this paper, implies that 
both constituent Ising lattices 
must have reached zero magnetization. Therefore, the lifetime of the clock 
model is the longer of the lifetimes of 
the two Ising models. If the applied field is in the $\vec e_2$ direction, 
the two Ising systems are identical, and the 
probability density of the lifetime for the clock model can be simply 
written as
\begin{equation}
\label{eq:cllt}
	p_{\rm c}(\tau) = \langle \tau_{\rm I} \rangle^{-1} 
	\exp(-\tau /2\langle \tau_{\rm I} \rangle)
               [1 - \exp(-\tau / 2\langle \tau_{\rm I} \rangle)]\; .
\end{equation}
The mean and relative standard deviation of this distribution are 
$\langle \tau \rangle = 3\langle \tau_{\rm I} \rangle$ and $r = \sqrt{5}/3$,
respectively. The derivation of $p_{\rm c}(\tau)$ for the general case 
that $\vec H$ is not parallel to $\vec e_2$ is detailed in Appendix B with 
the general result given as Eq.~(\ref{eq:pctau}).

\begin{figure}
\begin{center}
\vspace*{2.25in}
         \includegraphics{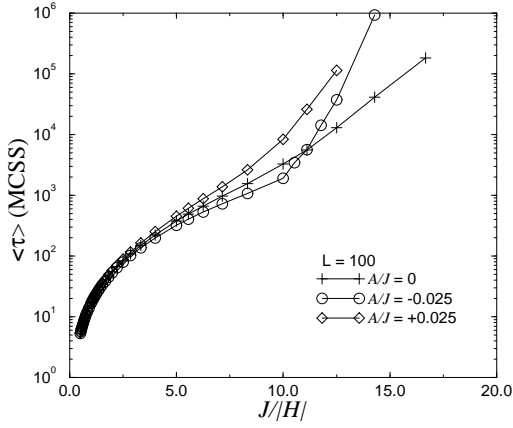}
\begin{picture}(300, 10)(0,0)
        \put(115, 10){(a)}
\end{picture}
\vspace{2.25in}
         \includegraphics{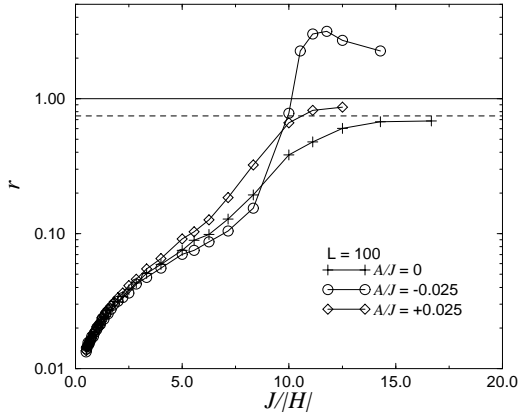}
\begin{picture}(300, 10)(0,0)
        \put(115, 10){(b)}
\end{picture}
\end{center}
\caption{$\langle \tau \rangle$ and $r$ $vs.$ $J/|H|$ for $L = 100$.}
\label{lt00}
\end{figure}

\begin{figure}
\begin{center}
\vspace*{2.25in}
         \includegraphics{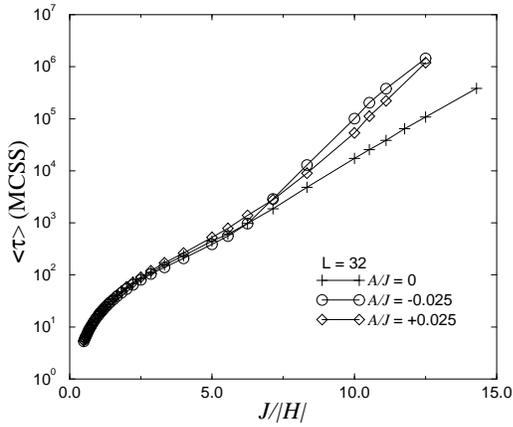}
\begin{picture}(300, 10)(0,0)
        \put(115, 10){(a)}
\end{picture}
\vspace*{2.25in}
         \includegraphics{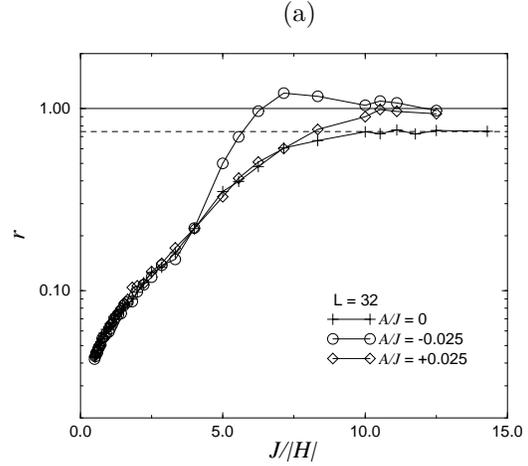}
\begin{picture}(300, 10)(0,0)
        \put(115, 10){(b)}
\end{picture}
\end{center}
\caption{$\langle \tau \rangle$ and $r$ $vs.$ $J/|H|$ for $L = 32$.}
\label{lt32}
\end{figure}

In the MD regime, since the switching process of the Ising model is 
deterministic, based on the mapping between the Ising model and the 
4-state clock model, the switching dynamics of the 4-state clock 
model is also deterministic. By ``deterministic'' we mean that 
the lifetime distribution is 
characterized by a very small relative standard deviation, 
$r \ll 1$. Figure~\ref{lt00} and Fig.~\ref{lt32} 
show $\langle \tau \rangle$ and 
$r$ as functions of the inverse applied field, $J/|H|$, for different lattice 
sizes. In these measurements, the applied field $\vec H$ is always parallel to 
$\vec{e_2}$. In the panels showing $r$, the horizontal solid and dashed lines
correspond to $r = 1$ and $r = \sqrt{5}/3$, respectively. As predicted, for 
the isotropic clock model,  $r$ approaches $\sqrt{5}/3$ as $|H|$ decreases. 
Also shown in these figures are results for the asymmetric clock model. 
When the easy axis is along $\vec{e_0}$--$\vec e_2$~($A>0$), the lifetime is 
longer than that of the symmetric model with the same applied field because a 
stronger field is needed to flip the spins from the metastable state 
$\vec e_0$ to the intermediate states, $\vec e_1$ and $\vec e_3$. When the 
easy axis is along $\vec{e_1}$--$\vec e_3$~($A<0$), however, 
the lifetime is shorter than that for the symmetric model 
in the MD regime, and longer in the SD regime.  The lifetime distribution 
is wider in the SD regime, and $r$ has a system size-dependent maximum for 
$A<0$. In either case, as $|H|$ decreases, the asymmetry dominates, and the 
two states along the easy axis are strongly favored over the other two 
states. Therefore for weak $\vec H$ we expect the asymmetric clock model to 
behave essentially 
like a single Ising model, so that $r$ will approach unity.

\begin{figure}
\begin{center}
\vspace*{2.8in}
	\includegraphics{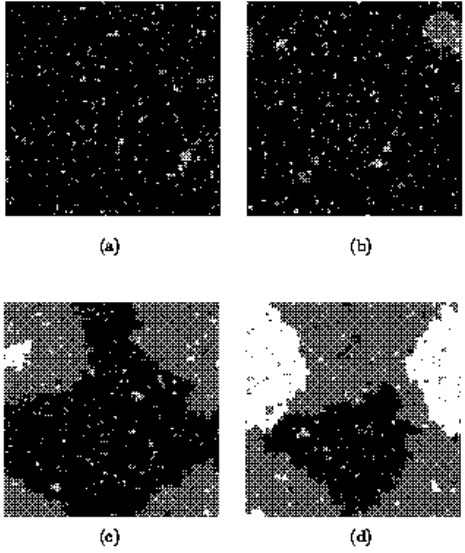}
\end{center}
\caption{One typical switching event for $A/J = 0.025$, $L=100$,
	$|H|/J = 0.09$. $\tau = 66000$~MCSS. 
	$\langle \tau \rangle = 25886$~MCSS, $r = 0.82$.
	The times for these four snap shots are 
	$t=$ 62010, 62681, 65130, and 65980~MCSS, respectively.}
\label{snap1}
\end{figure}

\begin{figure}
\begin{center}
\vspace*{2.8in}
	\includegraphics{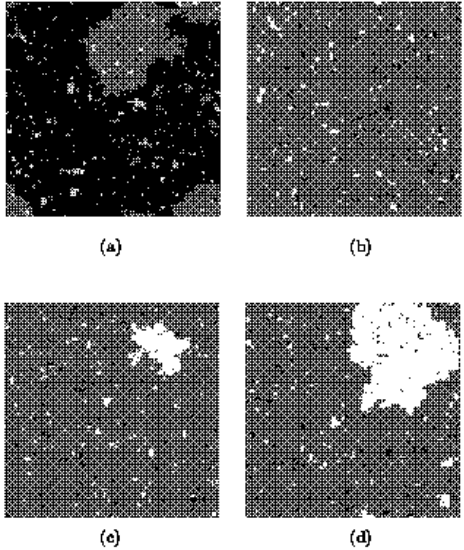}
\end{center}
\caption{One typical realization of decay via the slow channel for
	$A/J = -0.025$, $L=100$, $|H|/J = 0.09$. $\tau = 44542$~MCSS. 
	$\langle \tau \rangle = 5659$~MCSS, $r = 3.02$.
	The times for these four snap shots are 
	$t =$ 1100, 39960, 43130, and 44400~MCSS, respectively.}

\label{snap3}
\end{figure}

\begin{figure}
\begin{center}
\vspace*{2.8in}
	\includegraphics{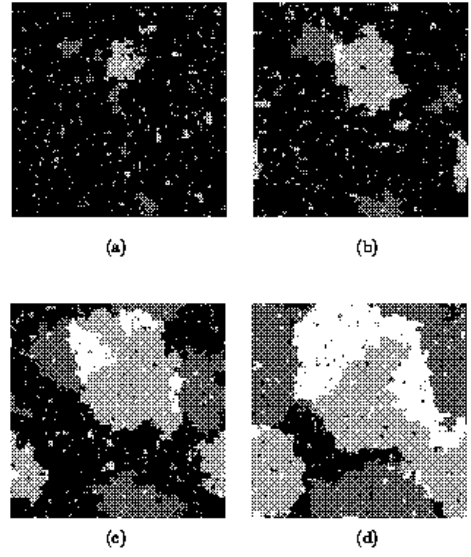}
\end{center}
\caption{One typical realization of decay via the fast channel for 
	$A/J = -0.025$, $L=100$, $|H|/J = 0.09$. $\tau = 1649$~MCSS. 
	$\langle \tau \rangle = 5659$~MCSS, $r = 3.02$.
	The times for these four snap shots are 
	$t =$ 410, 820, 1230, and 1640~MCSS, respectively.}
\label{snap4}
\end{figure}

In order to understand the switching process for the asymmetric model 
better, we show selected snapshots of the model during the simulations in 
Figs.~\ref{snap1}-\ref{snap4}. The darkest regions represent the initial 
state, the lightest ones represent the stable state, and the two 
intermediate shades of gray represent the intermediate states. The applied 
field is chosen so that $r$ is near its maximum for $A<0$~(Fig.\ref{lt00}).
For $A>0$ (Fig.~\ref{snap1}), there is 
only one channel for the model to switch the magnetization: first a droplet 
of one intermediate state is nucleated and grows quickly. Then a droplet of
the stable state nucleates and grows on this intermediate background. This
process is similar to single-droplet decay in the Ising model, except that 
there is no intermediate phase in the Ising model. For $A<0$, however, 
two channels exist. In the slow channel (Fig.~\ref{snap3}), the system 
first enters a uniform intermediate state, from which it takes a long 
time to nucleate a stable droplet. In the fast channel~(Fig.~\ref{snap4}), 
droplets in different intermediate states are formed simultaneously. As they 
meet, a stable droplet is quickly nucleated on their common boundary to 
decrease the free energy. Because of the existence of these two channels,
the relative standard deviation of the lifetime becomes very large, exceeding
unity, which is the maximum value for a single Ising model.

\begin{figure}
\begin{center}
\vspace*{2.2in}
         \includegraphics{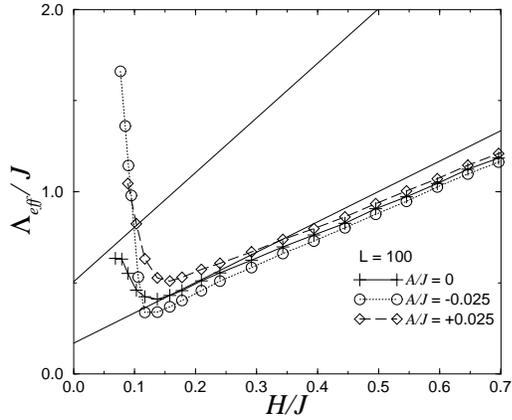}
\begin{picture}(300, 10)(0,0)
        \put(115, 10){(a)}
\end{picture}
\vspace*{2.2in}
         \includegraphics{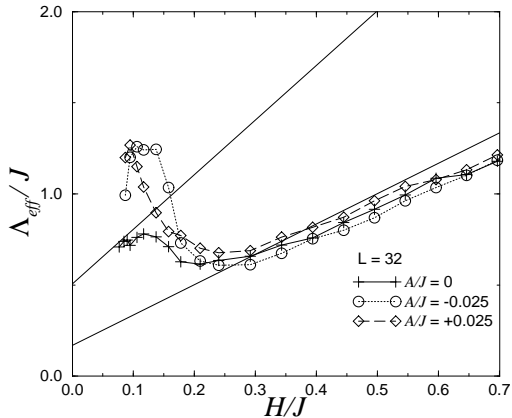}
\begin{picture}(300, 10)(0,0)
        \put(115, 10){(b)}
\end{picture}
\end{center}
\caption{$\Lambda_{\rm{eff}}/J$ $vs$. $|H|/J$. The solid lines are the 
theoretical predictions for the Ising model from Eq.~(\ref{eq:lamda}).}
\label{fig:dtime}
\end{figure}

Figure~\ref{fig:dtime} shows the derivatives of the logarithm of the lifetime 
with respect to $J/|H|$, $\Lambda_{\rm eff}$ of Eq.~(\ref{eq:lamda}), for 
different system sizes. In this figure, the straight lines are the 
theoretical results obtained from Eq.~(\ref{eq:lamda}). Indeed, for the 
symmetric clock model, our numerical results show good agreement with the
theory. Note that there are no adjustable parameters used in obtaining the
theoretical results.
In the asymmetric cases, when the applied field is strong the 
effect of the asymmetry is small. For weak fields, the asymmetry dominates 
the switching process, and the results for the clock model cannot be mapped 
onto those for the Ising model. Unfortunately, we could not extend the
simulations deep into this regime, due to the prohibitive computer resources 
required: it takes at least two months to simulate 1000 individual 
realizations for $L = 100$ and $J/|H| > 12.5$ (Fig.~\ref{lt00}) on a 
single node of an IBM sp2 computer.

\vfill
\eject

\section{Switching-Field Measurements}
\label{sec:switch}

We define the switching field as follows. We apply a field $\vec H$ 
to the system in the initial state for a chosen waiting time 
$t_{\rm w} = 200$~MCSS. If the magnetization has switched $N/2$ times 
in $N$ simulations, then the magnitude of the applied field is called the 
switching field, $H_{\rm sw}(t_{\rm w})$. We choose this definition in order
to simulate qualitatively the experiment discussed in Ref.~\cite{CHANG93}. The 
relationship between the switching field and the coercivity is that in
coercivity measurements, the applied field is changed 
sinusoidally or linearly~\cite{MEE86}, while in our simulation the applied 
field is reversed abruptly. 

\begin{figure}
\begin{center}
\vspace*{2.1in}
         \includegraphics{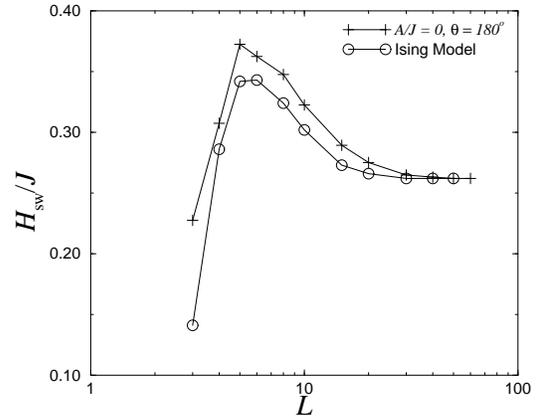}
\end{center}
\caption{Switching fields for the symmetric clock model and the Ising model.}
\label{fig-clkis}
\end{figure}

First we compare the switching field for the symmetric clock model and
the Ising model. The switching field as a function of the system size $L$ 
is illustrated in Fig.~\ref{fig-clkis}. Since the time scale of the clock 
model is twice that for the Ising model, the switching field for the Ising 
model is measured with waiting time $t_{\rm w} = 100$~MCSS.
In the MD regime, the lifetime for 
the symmetric clock model is twice that for the Ising model, 
thus the switching fields for the clock model and the Ising model are 
the same. In the stochastic regime, from Eq.~(\ref{eq:islt}) and 
Eq.~(\ref{eq:cllt}), the lifetime for the clock model is three times that of 
the Ising model. Therefore, the switching field for the clock model is larger
than that for the Ising model. Here we emphasize that the maximum 
shown in Fig.~\ref{fig-clkis} is the result of different switching mechanisms
dominating the decay in the CE, SD and MD regimes. It is {\it not} the 
effect of demagnetization since there are no dipole-dipole interactions in our 
Hamiltonian. For more detailed discussions of the dependence of $H_{\rm sw}$
on $L$, $t_{\rm w}$ and $T$ in the Ising model, see Ref.~\cite{RICH95}.

\begin{figure}[htb]
\begin{center}
\vspace*{2.1in}
         \includegraphics{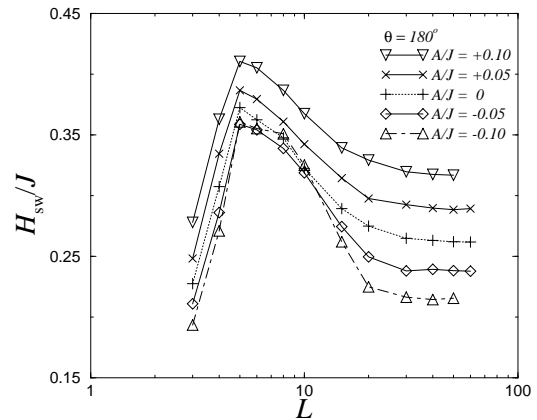}
\end{center}
\caption{Effect of $A$ on $H_{\rm{sw}}$ for $\theta = 180^\circ$.}
\label{anis}
\end{figure}

Next we consider the effects of the asymmetry $A$ on the switching 
field. The results are shown in Fig.~\ref{anis}. Again, $\vec H$ is in 
the $\vec{e_2}$ direction. When the easy axis is along 
$\vec e_0$--$\vec e_2$~($A>0$), increasing $A$ leads to a monotonic 
increase of the switching field. When $A<0$, on the other hand, making
$A$ more negative decreases the switching field in the MD regime. However, 
in the SD regime different $A$ shows little effect on the switching field.

\begin{figure}[htb]
\begin{center}
\vspace*{2.1in}
         \includegraphics{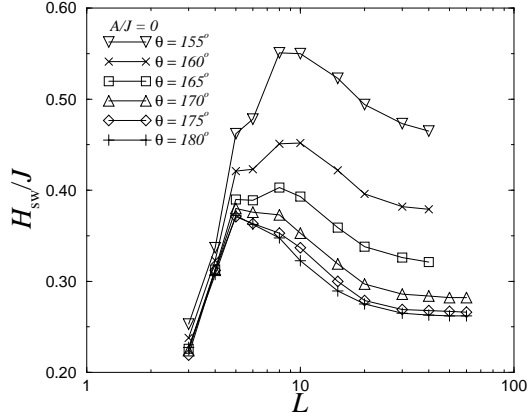}
\end{center}
\caption{Effect of $\theta$ on $H_{\rm{sw}}$ for $A = 0$.}
\label{angl}
\end{figure}

The effect of the direction of the applied field on the switching field is 
shown in Fig.~\ref{angl}. Here $H_{\rm{sw}}$ is shown as a function of $L$ 
for different $\theta$ at $A = 0$. The symmetric clock model is equivalent 
to two decoupled Ising models, and the field components acting on these two 
Ising systems change with $\theta$. Thus the switching behaviors of the 
two Ising systems are different, and the switching of the clock model 
is determined by the slower one. Therefore $H_{\rm sw}$ increases as $\vec H$ 
deviates further away from the $\vec{e_2}$ direction. 

\begin{figure}[htb]
\begin{center}
\vspace*{2.25in}
         \includegraphics{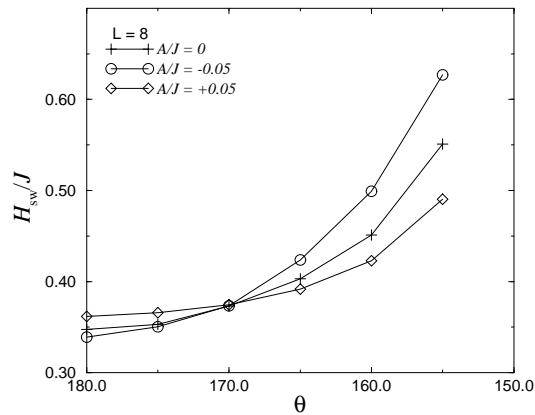}
\begin{picture}(300, 10)(0,0)
        \put(115, 10){(a)}
\end{picture}
\vspace*{2.25in}
         \includegraphics{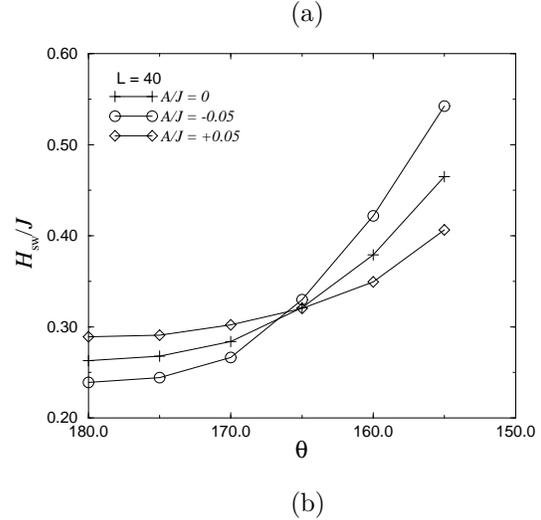}
\begin{picture}(300, 10)(0,0)
        \put(115, 10){(b)}
\end{picture}
\end{center}
\caption{$\theta$- and $A$- dependence of the switching field 
	for $L = 8$ and $40$.}
\label{anag}
\end{figure}

Finally we have measured the $\theta$-dependence of $H_{\rm{sw}}$ for two 
system sizes ($L = 8$ and $L = 40$) with different values of $A$. The results 
are shown
in Fig.~\ref{anag}. When the direction of $\vec H$ is close to $\vec e_2$, 
the energetic preference for the two intermediate states is similar and 
mainly determined by the asymmetry. Therefore $H_{\rm sw}$ is governed by $A$.
As $\theta$ approaches $135^\circ$, $\vec H$ increases the stability of one 
intermediate state and decreases that of the stable state. When $A < 0$, this 
tendency is further amplified, therefore the switching field 
increases. For $A>0$, $\vec H$ makes the flip from $\vec e_0$ to the favored
intermediate state easier, whereafter $A$ acts like an applied
field in flipping the spin to $\vec e_2$. Thus $H_{\rm{sw}}$ decreases 
compared with the isotropic model.

\section{Conclusion}
\label{sec:summ}
In this work we have studied magnetization switching in biaxial magnetic 
media by large-scale Monte Carlo simulation and theoretical analysis
of a kinetic 4-state clock model. The model reduces the infinite anisotropy in 
the kinetic Ising model by introducing two intermediate magnetization 
directions between the magnetization states parallel and antiparallel to 
the applied field. If the two perpendicular axes are equivalent, the 
switching process is equivalent to that of two decoupled Ising models. 
With asymmetry, the clock model shows a more complicated 
switching behavior. When the axis along the initial state
is favored ($A>0$), the model becomes Ising-like. On the other hand, if
the other easy axis is favored, the system can switch its magnetization via 
two different channels, which can lead to a wider distribution of the 
lifetimes in the SD regime. Another factor affecting the switching process 
is the direction of the applied field. We studied the direction dependence 
of the switching field in the isotropic model. With the applied field 
deviating further from the stable state, a stronger switching field is 
needed. The different switching mechanisms in the CE, SD, and MD regimes 
lead to a maximum switching field for a specific system size, in agreement with
experimental results. We conclude that the 4-state clock model
is a simple but informative platform to study the magnetization switching
in biaxial magnetic particles.

\section*{Acknowledgment}
The authors wish to thank Dr.~Howard~L.~Richards for proposing this study. 
We also thank Dr.~S.~M.~Durbin, Dr.~S.~W.~Sides, Dr.~G.~Korniss, 
Dr.~H.~L.~Richards, Dr.~M.~Kolesik and Dr.~G.~Brown for useful comments.
This work is supported in part by the US National Science Foundation 
Grant No.\ DMR-9520325, and by Florida State University through the
Supercomputer Computations Research Institute (US Department of Energy
Contract No.\ DE-FC05-85ER25000) and the Center for Materials Research and 
Technology.

\section*{Appendix A}
In this appendix, the dynamical mapping between the Ising model and the 
symmetric 4-state clock model is derived. 

\begin{figure}[hbt]
\begin{center}
	\vspace*{1.3in}
	\includegraphics{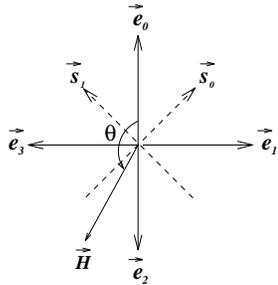} 
\end{center}
\caption{Mapping of the 4-state clock model onto a superposition of
	two Ising models.}
\label{fig:clkvc}
\end{figure}

First we define two unit vectors, $\vec s_0$ and $\vec s_1$, such that the 
angle between $\vec s_0$ ($\vec s_1$) and $\vec e_0$ is 
$-45^\circ$~($+45^\circ$) (Fig.~\ref{fig:clkvc}). Let $\vec \mu_i$ denote the 
direction of the clock spin at site $i$, which can be $\vec e_0$,
$\vec e_1$, $\vec e_2$, or $\vec e_3$. The unit vector $\vec \mu_i$ can be 
expressed in terms of $\vec s_0$ and $\vec s_1$ as 
\begin{equation}
  \label{SPMAP}
        \vec \mu_i = \frac{1}{\sqrt{2}}(\omega_{0i} \vec s_0 + 
		\omega_{1i}  \vec s_1)\; ,
\end{equation}
where $\omega_{0i}$ and $\omega_{1i}$ can take the values $\pm 1$. The 
applied field 
can also be decomposed as
\begin{equation}
  \label{FIELD}
        \vec H = H_0 \vec s_0 + H_1 \vec s_1\; ,
\end{equation}
where
\begin{eqnarray}
  \label{fdcom0}
H_0 & = & |\vec H|\cos (45^\circ + \theta)\; , \\
%  \label{fdcom1}
H_1 & = & |\vec H|\sin (45^\circ + \theta)\; .
\end{eqnarray}

Substituting Eqs.~(\ref{SPMAP})-(\ref{fdcom0}) into Eq.~(\ref{eq:CLKHM}), 
we get
\begin{eqnarray}
\label{fnhmt}
  {\cal H} & = & {\cal H}_0 + {\cal H}_1\; , \\
\label{ishm0}
  {\cal H}_0 & = & -\frac{J}{2} \sum_{\langle i,j \rangle} \omega_{0i} \omega_{0j}
             -\frac{H_0}{\sqrt{2}}\sum_i \omega_{0i}\; , \\
\label{ishm1}
  {\cal H}_1 & = & -\frac{J}{2} \sum_{\langle i,j \rangle} \omega_{1i} \omega_{1j}
             -\frac{H_1}{\sqrt{2}}\sum_i \omega_{1i}\; .
\end{eqnarray}
From these equations, it is clear that the Hamiltonian of the symmetric
4-state clock model is equivalent to the sum of two decoupled Ising 
Hamiltonians with exchange interactions $J/2$.
The applied field in each Ising Hamiltonian equals the field
component times $1/\sqrt{2}$. Therefore, the 4-state clock model in equilibrium
can be treated as two superimposed Ising systems, with two Ising spins at
each site~\cite{MASUO}.

The next step is to construct a mapping of the spin-flip dynamics 
between the 4-state clock model and the Ising models.
Since each spin can only flip to its neighboring state, this can be done by 
choosing one of the two Ising systems randomly and flipping the corresponding 
Ising spin. Because each time only one Ising lattice is 
selected, the change of the energy due to the spin flip
is obtained by updating either ${\cal H}_0$ or ${\cal H}_1$ in 
Eq.~(\ref{fnhmt}).

With the above analysis, it follows that the dynamics of the symmetric 4-state
clock model is equivalent to that of two decoupled Ising models. Not only
are the spin-flip mechanisms interchangeable, but also 
only one Ising Hamiltonian in Eq.~(\ref{fnhmt}) is affected in each flip 
because
${\cal H}_0$ and ${\cal H}_1$ are decoupled. Since each clock spin
can be thought to be composed of two Ising spins, during the simulation 
the time scale for the 4-state clock model is twice that for the Ising model. 
The mapping of the parameters between the symmetric 4-state clock model and 
Ising models is summarized in Table~\ref{mapping}. In this table, $t$ 
is the time measured in Monte Carlo steps per spin (MCSS).
\begin{table}
  \caption{Mapping between the symmetric 4-state Clock and Ising Models}
  \label{mapping}
\begin{center}
  \begin{tabular}{|c|c|c|} \hline
    	 Clock & Ising system \#0 & Ising system \#1	\\ \hline \hline
        $J$   & $J/2$         & $J/2$		\\ \hline
     $\vec H$ & $\frac{1}{\sqrt{2}} |\vec H|\cos(45^\circ + \theta)$ 
              & $\frac{1}{\sqrt{2}} |\vec H|\sin(45^\circ + \theta)$ \\ \hline
        $t$             & $t/2$          & $t/2$\\ \hline
  \end{tabular}
\end{center}
\end{table}

\section*{Appendix B}

In this appendix, the lifetime distribution for the isotropic 4-state clock 
model is derived. We denote the lifetimes of the two Ising models by $\tau_0$
and $\tau_1$, and the lifetime of the 4-state clock model by $\tau_{\rm c}$. 
With the stopping criterion used in this work, we have
\begin{equation}
	\tau_{\rm c} = 2 \max(\tau_0, \tau_1)\; .
\end{equation}
Thus, for a given value $\tau$ of the lifetime,
the probability that $\tau_{\rm c} < \tau$ can be written as
\begin{equation}
\label{CDF}
	P(\tau_{\rm c}<\tau) = P(\max(\tau_0, \tau_1)<\tau/2)\; .
\end{equation}
Here the factor 2 comes from the mapping of the time scale between the
clock and the Ising models given in Table~\ref{mapping}. Since the two Ising 
models are decoupled, the lifetime distributions are independent. Therefore
\begin{equation}
	\label{eq:prod}
	P(\tau_{\rm c}<\tau) = P(\tau_0<\tau/2) \cdot P(\tau_1<\tau/2)\; .
\end{equation}

In the SD regime, the lifetime of each Ising model is exponentially 
distributed,
\begin{equation}
\label{eq:logtime}
	P(\tau_k < \tau) = 1-\exp(-\tau / \langle \tau_k \rangle)\; ,
\end{equation}
where $k=0,1$. Substituting Eq.~(\ref{eq:logtime}) into Eq.~(\ref{eq:prod}) 
and differentiating with respect to $\tau$, we obtain the
probability density of
the lifetime of the clock model as
\begin{eqnarray}
\label{eq:pctau}
p_{\rm c}(\tau)& = &\frac{\exp(-\tau/2\langle\tau_0\rangle)
	    \left[1-\exp(-\tau/2\langle \tau_1 \rangle)\right]}
	    {2\langle \tau_0 \rangle} \nonumber \\
	& & + \frac{\exp(-\tau/2\langle\tau_1\rangle)
	    \left[1-\exp(-\tau/2\langle \tau_0 \rangle)\right]}
	    {2\langle \tau_1 \rangle}\; .
\end{eqnarray}
If $\langle\tau_0\rangle = \langle\tau_1\rangle = \langle\tau_{\rm I}\rangle$,
we get Eq.~(\ref{eq:cllt}).

In the MD regime, since the magnetic switching is deterministic, the lifetime
probability density of the Ising model is well approximated by a 
$\delta$-function. Thus for the Ising model, 
\begin{equation}
	P(\tau_{\rm I} < \tau) \approx \eta(\tau - \tau_{\rm I})\; ,
\end{equation}
where $\eta$ is the unit step function. Following the same procedure as above
we get the lifetime distribution for the clock model,
\begin{equation}
	p_{\rm c}(\tau) \approx \delta(\tau - 2\max[\langle \tau_0 \rangle, 
			\langle \tau_1 \rangle])\; .
\end{equation}

Finally we consider the case that the field is applied in such a direction 
that one Ising system is in the MD
regime with average lifetime $\langle \tau_{\rm MD} \rangle $, and the other 
in the SD region with the average lifetime $\langle \tau_{\rm SD} \rangle$. 
Starting from Eq.~(\ref{eq:prod}), we then have
\begin{equation}
	 P_{\rm c}(\tau_{\rm c} < \tau) = \eta(\tau- 2\tau_{\rm MD})
		[1-\exp(-\tau / 
2\langle \tau_{\rm SD} \rangle)]\; , 
\end{equation}
which yields
\begin{eqnarray}
	p_{\rm c}(\tau)& = &[ 1 - \exp(-\langle \tau_{\rm MD} \rangle
				      /2\langle \tau_{\rm SD} \rangle)
			  ] \delta(\tau - 2\langle \tau_{\rm MD} \rangle)
				\nonumber \\
		& & +\frac{\exp(-\tau/2\langle \tau_{\rm SD} \rangle)}
				{2 \langle \tau_{\rm SD} \rangle}
				\eta(\tau - 2\langle \tau_{\rm MD} \rangle)\; .
\end{eqnarray}

A more detailed calculation for the MD cases would replace the 
$\delta$-functions by Gaussians and
the step functions by the corresponding error functions~\cite{RICH95},
with no essential change from the approximate results given here.

%%%%%%%%%%%%%%%%% BIBLIOGRAPHY IN THE LaTeX file !!!!! %%%%%%%%%%%%%%%%%%%%%%%%
%% This is nothing else than the IEEEsample.bbl file that you would          %%
%% obtain with BibTeX: you do not need to send around the *.bbl file         %%
%%---------------------------------------------------------------------------%%
%

\end{document}